\documentclass[a4paper,nobibnotes,nofootinbib,preprintnumbers]{revtex4}
\usepackage{amssymb}
\usepackage{amsmath}
\usepackage{epsfig}
\newcommand{\be}{\begin{equation}}
\newcommand{\ee}{\end{equation}}
\newcommand{\bea}{\begin{eqnarray}}
\newcommand{\eea}{\end{eqnarray}}

\newcommand{\bi}{\begin{itemize}}
\newcommand{\ei}{\end{itemize}}

\begin{document}
\title{Geometric Scaling of $F_2$ and $F_2^c$ in data and QCD Parametrisations}
\author{Guillaume Beuf}\email{guillaume.beuf@cea.fr}
\affiliation{Institut de physique th{\'e}orique, CEA/Saclay, 91191
Gif-sur-Yvette cedex, France
\\URA 2306, unit{\'e} de recherche associ{\'e}e au CNRS}
\author{Christophe Royon}\email{christophe.royon@cea.fr}
\affiliation{IRFU/Service de Physique des Particules, CEA/Saclay, 91191
Gif-sur-Yvette cedex, France}
\author{David \v S\'alek}\email{salekd@mail.desy.de}
\affiliation{Institute of Particle and Nuclear Physics, Charles University, Prague,
Czech Republic}


\begin{abstract}
The scaling properties at low $x$ of the proton DIS cross section  and 
its charm component are analyzed with the help of the quality factor
method. Scaling properties are tested both in the deep inelastic scattering
data and in the structure functions reconstructed from CTEQ, MRST and GRV
parametrisations of parton density functions. The results for DIS cross
sections are fully
compatible between data and parametrisations. Even with larger error bars, the
charm component data favors the same geometric scaling properties as the ones of
inclusive DIS.
This is not the case for all parametrisations of the charm component.
\end{abstract}
\maketitle
\section{Introduction}

Geometric scaling \cite{Stasto:2000er} is a remarkable empirical
property
verified by data on high energy deep inelastic scattering (DIS) $i.e.$
virtual photon-proton cross-sections.
One can indeed represent with
reasonable
accuracy the cross section $\sigma^{\gamma^*p}$ in the low Bjorken $x$ regime by the formula
\begin{equation}
\sigma^{\gamma^*p}(x,Q)=\sigma^{\gamma^*}(\tau)\  ,
\quad \textrm{where} \quad
\tau = \log \left( \frac{Q^2}{Q_s^2(Y)}\right)
\label{tau}
\end{equation}
is the scaling variable,
$Q$ the virtuality of the photon, and
$Y\equiv \log(1/x)$ the total rapidity in the ${\gamma^*}$-proton system.
That empirical property is often considered as a hint for gluon saturation effects.
Indeed, including non-linear effects due to high parton density in the evolution equations for parton distributions towards low $x$ leads to such geometric scaling properties. In that context, $Q_s(Y)$ is called the saturation scale, and gives the typical scale for the onset of nonlinear effects. 
Using the dipole factorization of DIS observables, one predicts geometric 
scaling to hold also for deeply virtual Compton scattering, exclusive vector 
meson production and inclusive diffraction, which has been verified 
empirically~\cite{Marquet:2006jb}.
Several types of geometric scaling or of generalizations of geometric scaling have been proposed in the literature, which correponds to various sets of approximations in the theoretical description of parton evolution and gluon saturation.

The quality factor (QF) introduced in \cite{usgeom} is an interesting tool in 
order to discuss scaling properties. It allows to compare the validity of 
different scaling laws on a given data set, without any assumption 
concerning 
the shape of the scaling function. It has been used in \cite{usgeom,usscaling} 
in order to study the above-mentioned scaling properties in the data for DIS 
observables.
A standard fit to data tests locally in each bin the value of an observable, 
whereas the QF only considers point-to-point correlations for the observable. 
Hence, the QF method could give complementary informations, difficult to catch 
with a standard fit if the data are not very precise. For that reason, it is 
interesting to test with the QF method to what extend global fits to parton 
distribution functions (PDF) leads to the same scaling properties as the data. 
Many efforts have been made in the last few years by the MRST and CTEQ 
groups 
in order to implement heavy quark mass effects in a consistent way in their 
global fits. 
That justifies the comparison of data and PDF sets not only for the inclusive
DIS cross section
$\sigma \sim F_2/Q^2$ but also for its charm component $
\sigma^c \sim F_2^c/Q^2$. The data for the bottom 
component $F_2^b/Q^2$ are not suitable for a scaling analysis, since they contain 
too few points.

The plan of the paper is as follows. In section {\bf II} we review recent improvements in the treatment of heavy flavors in PDF's global fits, and we present the PDF sets used in the present study. In section
{\bf III} we describe the various scaling hypotheses to be tested and their theoretical motivation. In section {\bf IV} we present the
Quality Factor method we shall use in the analysis and the set of data we
consider for the fitting procedure. We provide and comment our results for $F_2$ in
section {\bf V} and for $F_2^c$ in
section {\bf VI}. The section {\bf VII} is devoted to the summary and the discussion of the results.


\section{PDFs with heavy quark mass effects}\label{sec:PDFs}

Let us briefly discuss the recent progresses concerning the treatment of heavy 
quark mass effects in PDFs global fits\footnote{For a recent pedagogical 
review, see \cite{Thorne:2008xf}.}. In the collinear factorization framework, 
one factorizes observables (\emph{e.g.} proton structure functions) as a 
convolution of coefficient functions, describing partonic matrix elements, and 
PDFs for active flavors. In that formalism, active flavors are massless. 
Thus it should be relevant only if the quark masses corresponding to active 
flavors are small compared to the factorization scale $\mu_F$, usually chosen 
to be a hard scale of the problem, such as $Q$ for DIS. The most standard 
pQCD scheme for heavy flavors is the \emph{fixed flavor number scheme} (FFNS), 
where heavy flavors are not considered as active, and are generated only by 
boson-gluon fusion. It gives reliable results for $Q=\mu_F$ of the order of 
the heavy quark mass $m_{hf}$. However, for $Q^2\gg m_{hf}^2$ higher order 
terms in the coefficient functions are enhanced by powers of 
$\log(Q^2/m_{hf}^2)$, so that perturbation theory breaks down. In order to 
overcome this difficulty, \emph{variable flavor number schemes} (VFNS) have 
been introduced. They consist in a succession of FFNS with one additional 
active flavor each time $Q$ becomes larger than the mass of a heavy quark 
flavor. Two successive FFNS are matched with each other at the value 
$Q^2\simeq m_{hf}^2$ chosen for the change of scheme, in order to ensure that 
renormalization is done in a consistent way for all FFNS. In the the first 
implementations of VFNS, called \emph{zero-mass}-VFNS (ZM-VFNS), the 
switching from one FFNS to another at $Q^2\simeq m_{hf}^2$, which is an 
unphysical change of factorization scheme, was coinciding with the threshold 
of production of the heavy flavor in the final state. That is not considered 
to be correct, and the latter should correspond to a threshold in the total 
energy $W$ rather than in $Q$. That difficulty and a few others are solved if 
one goes to the \emph{general-mass}-VFNS (GM-VFNS), build in \cite{Collins}. 
One of the ingredients used in the GM-VFNS is the replacement of $x$ by the 
rescaled variable
\begin{equation}
\chi=x \left(1+\frac{4 m_{hf}^2}{Q^2}\right)\label{ACOTchi}
\end{equation}
when a pair of heavy quarks $H \bar{H}$ is produced in the final state, in 
order to ensure the right kinematical threshold: for $W$ as small as 
$2 m_{hf}$, $\chi$ goes to $1$ and thus the cross section for heavy quark pair 
production vanishes.

The recent global fits of PDFs of the CTEQ and MRST groups are using such 
GM-VFNS. However, the details of the implementation differs between the two 
groups \cite{Tung:2006tb,Thorne:2006qt}. In particular, they use different 
conventions to name the order of the analysis 
(see \emph{e.g.} \cite{Thorne:2008xf}), such as LO, NLO, etc.
In our study, we use the following general purpose GM-VFNS PDFs sets: the NLO 
CTEQ6.6M \cite{Nadolsky:2008zw}, the NNLO MRST2006 \cite{Martin:2007bv}, and 
the NLO MRST2004 \cite{Martin:2004ir}. We consider also the NNLO 
MRST2004 \cite{Martin:2004ir}, which only uses an approximate implementation 
of GM-VFNS, and the older GRV98 \cite{Gluck:1998xa} PDFs set based on a FFNS 
with 3 flavors.

In all of the abovementioned parametrisations, the charm production in DIS 
appears only pertubatively and the charm and anticharm PDFs (if any) start 
from zero at an intial factorization scale below the charm mass. 
Alternatively, some theoretical models for the nucleon content feature 
a non-perturbative intrinsic charm component. The CTEQ group studied that 
possibility, and released the PDFs sets CTEQ6.6C1 to 
CTEQ6.6C4 \cite{Nadolsky:2008zw}, which extend CTEQ6.6M by including various 
types of intrinsic charm component. The CTEQ6.6C1 and CTEQ6.6C2 
parametrisations correspond respectively to a moderate and a strong intrinsic 
charm contribution of the form predicted by the BHPS 
model \cite{Brodsky:1980pb}, which is non-negligible only at large x. 
By contrast, the CTEQ6.6C3 and CTEQ6.6C4 parametrisations correspond 
respectively to a moderate and a strong intrinsic charm contribution 
proportional to the light sea quark contribution at the initial scale for the 
global fit. They are not motivated by theoretical models.


\section{Scaling Variables from saturation physics}

Let us sketch the theoretical motivation for the different forms of scaling 
(see TABLE I) proposed for deep-inelastic scattering at high energy. 
The evolution of unintegrated parton densities towards large rapidity $Y$ at 
fixed $k_T^2$ is given by the BFKL \cite{Lipatov:1976zz} equation, provided 
that these unintegrated parton densities are small. Generic solutions of that 
linear evolution can be written as a sum of elementary wave solutions, with 
the help of Mellin transform. Each of these elementary wave solutions have 
scaling properties, but generically not their superposition. However, the BFKL 
evolution in $Y$ leads to larger and larger unintegrated gluon density due to 
soft gluon radiation. At some point, the partons are so packed that additional 
soft gluons are emitted collectively. Thus, the soft gluons radiation is 
reduced by destructive interferences and the evolution becomes nonlinear in 
domain typically given by $k_T < Q_s(Y)$. The BFKL equation is then 
generalized \emph{e.g.} by the BK \cite{Balitsky:1995ub} or 
JIMWLK \cite{JIMWLK} nonlinear equations. For a review about that gluon 
saturation phenomenon, see \cite{Iancu:2003xm}. Such a nonlinear evolution 
equation has the following property. Even in the kinematical domain in which 
the nonlinear terms are negligible compared to the linear ones, the 
nonlinearity of the equation constrains the solution, acting as a dynamical 
boundary condition to the linear BFKL evolution \cite{Mueller:2002zm}. 
The interplay between that specific boundary condition and the BFKL kernel 
selects dynamically \cite{Munier:2003vc} a specific wave solution 
(the \emph{critical} one) of the BFKL equation in a window 
$k_T^2\gtrsim Q_s(Y)$, which determines the evolution of $Q_s(Y)$.
In that kinematical range, the solution of the nonlinear evolution loses memory of the initial condition. Hence, one finds that the unintegrated gluon distribution scales with $k_T^2/Q_s(Y)$ above the saturation scale. Due to the $k_T$-factorization in the linear regime, that property result in a geometric scaling \eqref{tau} of the DIS cross-section, and of other observables.

In a fixed coupling $\alpha_s$ approximation, that mechanism is well 
understood, and leads to a saturation scale $\log Q_s(Y) \sim \lambda Y$. 
This corresponds to the geometric scaling variable FC in TABLE I.
If one goes to the running coupling case, the mechanism leading to the scaling 
holds, but is analytically under control only at large enough $Q^2$, due to 
asymptotic freedom. One can deduce only the large $Y$ behavior of the 
saturation scale, which is $\log Q_s(Y) \sim \lambda \sqrt{Y}$. Thus, one gets 
the predicted scaling property at large $Y$ and $Q^2$ (with 
$\log Q^2 \propto \sqrt{Y}$). However, the extrapolation of that scaling 
property towards the finite $Q^2$ and $Y$ phenomenologically relevant domain 
is not unique. 
We consider in particular two different scaling variables, which both give the 
theoretical asymptotic behavior. 
The first one is a genuine geometric scaling variable 
(as defined in \eqref{tau}). It is called RCI in TABLE I. The second scaling 
variable \cite{gb} compatible with the asymptotic prediction is given in 
TABLE I and called RCII. Its value for the parameter $\lambda$ should be the 
square of the one for the RCI scaling.
In the recent years, the possible impact on saturation of fluctuations 
due \emph{e.g.} to Pomeron loops has been extensively discussed. They lead to 
a random saturation scale, but event-by-event the geometric scaling property 
is preserved. In the fixed coupling approximation\footnote{In the running coupling case, such a scenario is expected 
to hold in the high $Y$ limit \cite{Beuf:2007qa}. However, the running of the 
coupling seems to suppress the fluctuations effects in the accessible rapidity 
range at collider experiments \cite{Dumitru:2007ew}.}, the distribution of 
$\log Q_s^2(Y)$ is approximately gaussian, with a variance proportionnal 
to $Y$. Finally, one expect the cross section to scale with the diffusive 
scaling variable \cite{Hatta:2006hs} (DS in TABLE I) which is the original 
geometric scaling variable divided by the dispersion 
$\sqrt{Y}$.

\section{The Quality factor method}

In this section, we remind briefly the definition of the Quality Factor
($QF$) which we use to compare quantitatively the
different scaling variables (see Table~\ref{scalings}).
We already used this method
to compare the scaling results for the proton structure
function $F_2$, the deeply virtual Compton scattering (DVCS), the diffractive structure
function, and the exclusive vector meson production data measured at HERA~\cite{usscaling} .

Given a set of data points $(Q^2, x, \sigma=\sigma(Q^2,x))$ and a parametric
scaling variable $\tau = \tau(Q^2, Y=\log(1/x); \lambda)$ we want to know
whether the cross-section can be parametrised as a function of the variable $\tau$ only.
Since the function of $\tau$ that describes the data is not known,
the $QF$ has to be defined independently of the form of that function.
The $QF$ allows to quantitatively describe
whether pairs of $\sigma=\sigma(Q^2,x)$ and $\tau = \tau(Q^2, Y=\log(1/x); \lambda)$
lie on a single curve for a given parameter $\lambda$.

For a set of points $(u_i, v_i)$, where $u_i$'s are ordered,
we introduce $QF$ as follows~\cite{usgeom}

\be
QF(\lambda) = \biggl[ \sum_{i} \frac{(v_i-v_{i-1})^2}{(u_i-u_{i-1})^2+\epsilon^2} \biggr]^{-1}
\label{QF},
\ee
where $\epsilon$ is a small constant that prevents the sum from being infinite in case
of two points have the same value of $u$.
According to this definition, the contribution to the sum in~\eqref{QF} is large
when two successive points are close in $u$ and far in $v$.
Therefore, a set of points lying close to a unique curve
is expected to have larger $QF$ (smaller sum in~\eqref{QF})
compared to a situation where the points are more scattered.

Since the cross-section in data differs by orders of magnitude and $\tau$ is
more or less linear in $\log(Q^2)$ (see Table~\ref{scalings}),
we decided to take $u_i = \tau_i(\lambda)$ and $v_i = \log(\sigma_i)$.
This ensures that low $Q^2$ data points contribute to the $QF$
with a similar weight as higher $Q^2$ data points.
The set $(u_i, v_i)$ has to be ordered in $u$ before entering the $QF$ formula.
In order to stay independent of the exact values of $u$ and $v$,
all the values are rescaled so that $0 \le u_i,v_i \le 1$.
All the $QF$'s in this Letter are calculated with $\epsilon=0.01$.

In order to test a scaling law $\tau$ and say whether the points $(\sigma_i, \tau_i)$
lie on a single line or not,
we search for the parameter $\lambda$ that minimises the $1/QF$ variable.
The minimum value of $1/QF$ is obtained using the \verb+MINUIT+ package.
Given the maximum value of the $QF$ (minimum of $1/QF$),
we are able to directly compare different scaling laws.

\begin{table}
\begin{center}
\begin{tabular}{|c||c||c|c|} \hline
   & scaling & $\tau$ formula & parameters \\
\hline\hline
FC & ``Fixed Coupling" & $\log Q^2 - \lambda Y$ & $\lambda$ \\
\hline
RCI & ``Running Coupling I" & $\log Q^2 - \lambda \sqrt{Y}$ & $\lambda$ \\ \hline
RCII & ``Running Coupling II" &  $\log (Q^2/\Lambda^2) - \lambda
\frac{Y}{\log (Q^2/\Lambda^2)}$& $\lambda$ \\
 & & & $\Lambda=$0.2 GeV \\ \hline
DS & ``Diffusive Scaling" & $\frac{\log (Q^2/\Lambda^2) - \lambda Y}{\sqrt{Y}}$ & $\lambda$ \\
& & & $\Lambda=$1 GeV \\ \hline
\end{tabular}
\end{center}
\caption{Scaling variables used in the fits to deep inelastic scattering
data~\cite{usscaling}.}
\label{scalings}
\end{table}

In this Letter, we aim to extend the study performed in Ref.~\cite{usscaling} by
considering the different $F_2$ parametrisations and the proton charm structure
function data $F_2^c$ measured at HERA and given by different parametrisations.


\section{Fits to inclusive DIS}

We first aim to test the scaling quality using parametrisations
of the proton structure function $F_2$ using the 
parametrisations from the CTEQ, MRSTW and
GRV groups discussed in section \ref{sec:PDFs}.

In a previous paper~\cite{usscaling}, we tested the
scaling properties of $F_2/Q^2$ using the data available from the
H1~\cite{H1}, ZEUS~\cite{ZEUS}, NMC~\cite{NMC} and E665~\cite{E665} experiments.
We follow the same kinematical cuts for the parametrisations as we used for the experimental
data~\cite{usscaling}. We only consider data with $x<10^{-2}$
and $Q^2$ in the range $[3; 150]$~GeV$^2$.
The points with $x>10^{-2}$ are removed from the data since
valence quark densities dominate here,
and the formalism of saturation cannot apply in this kinematical region.
Similarly, the upper $Q^2$ cut is introduced,
while the lower $Q^2$ cut ensures that we stay away from the soft QCD domain.
In this kinematical domain, 217 data points are used.
In order to be able to directly compare scaling properties
in the CTEQ, MRST and GRV parametrisations to the experimental data,
we choose the same 217 pairs of $x$ and $Q^2$ as in data.

The $QF$ method is more likely to prefer the parametrisations over data,
since there is no statistical fluctuation in the parametrisations, contrary to
data.
It becomes clear from the definition that the statistically scattered values
of structure functions in the experimental data lead to worse $QF$
than a smooth parametrisation.
In this sense, we cannot use the value of the $QF$ to compare
the scaling properties between experimental data and parametrisations.
However, the $QF$ can be directly compared among different
parametrisations and can also test
the differences among scaling laws tested on
one particular data set.

\begin{figure}[t]
\begin{center}
\begin{tabular}{cc}
\epsfig{file=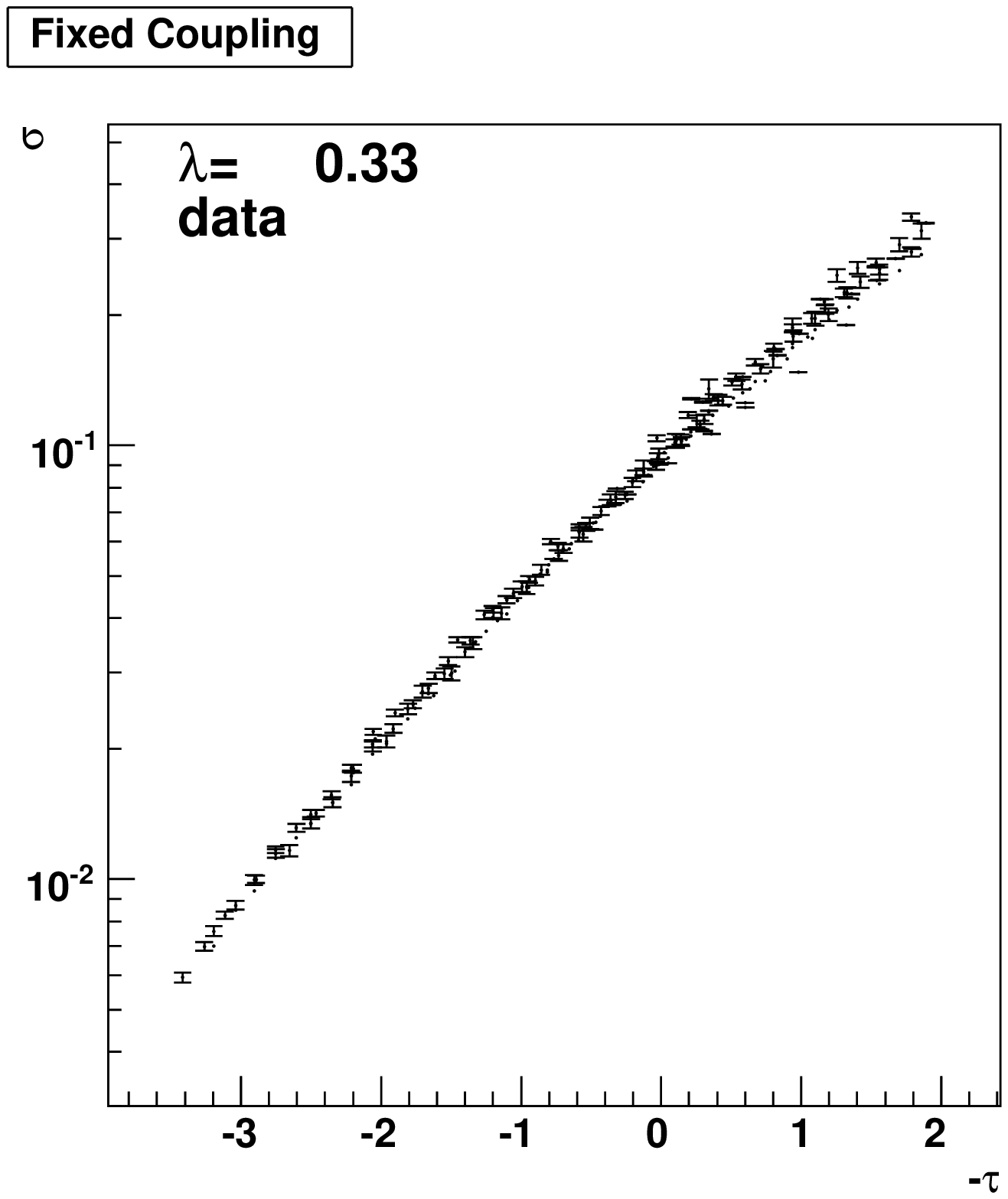,width=9.cm} &
\epsfig{file=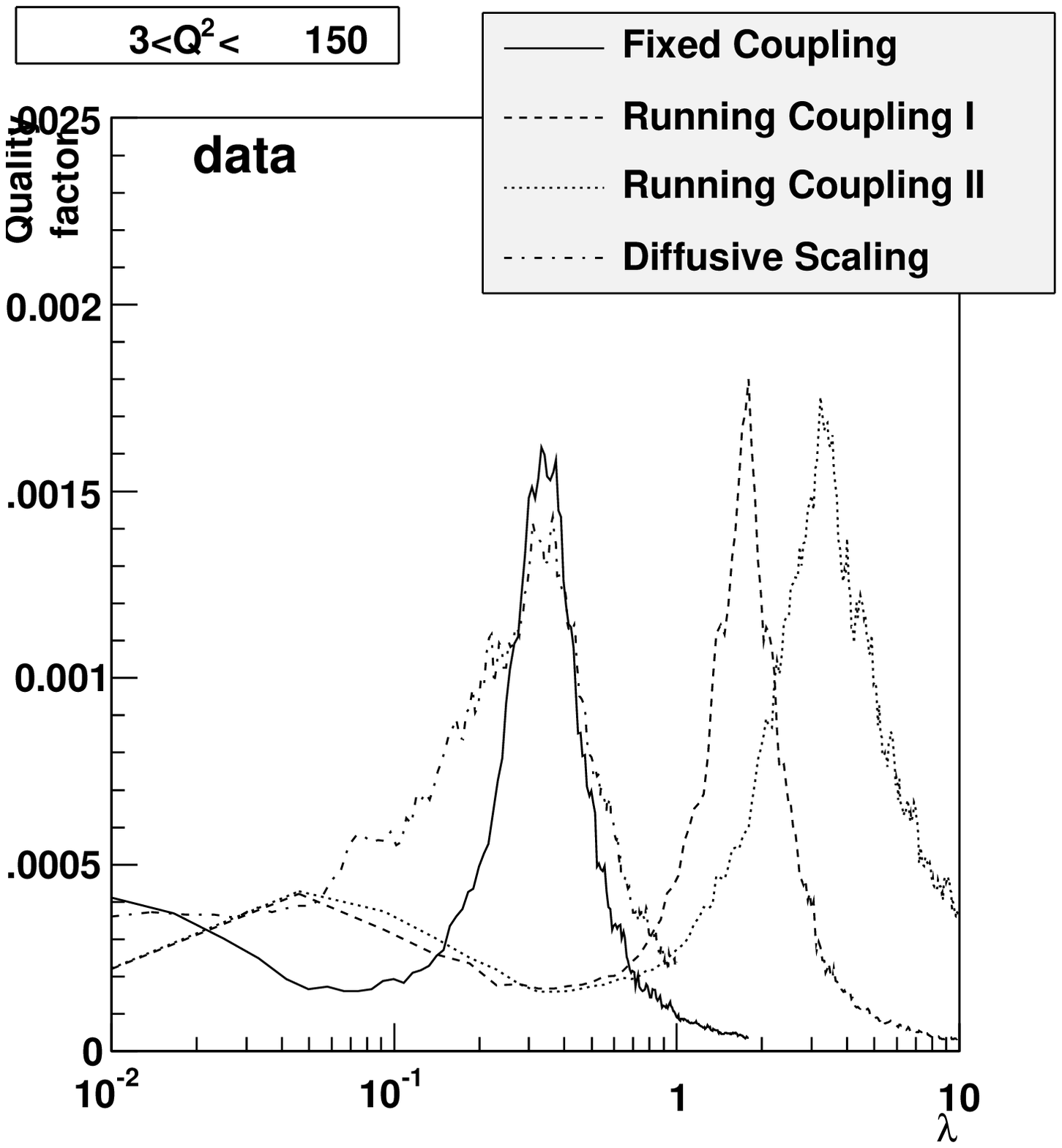,width=9.cm} \\
\epsfig{file=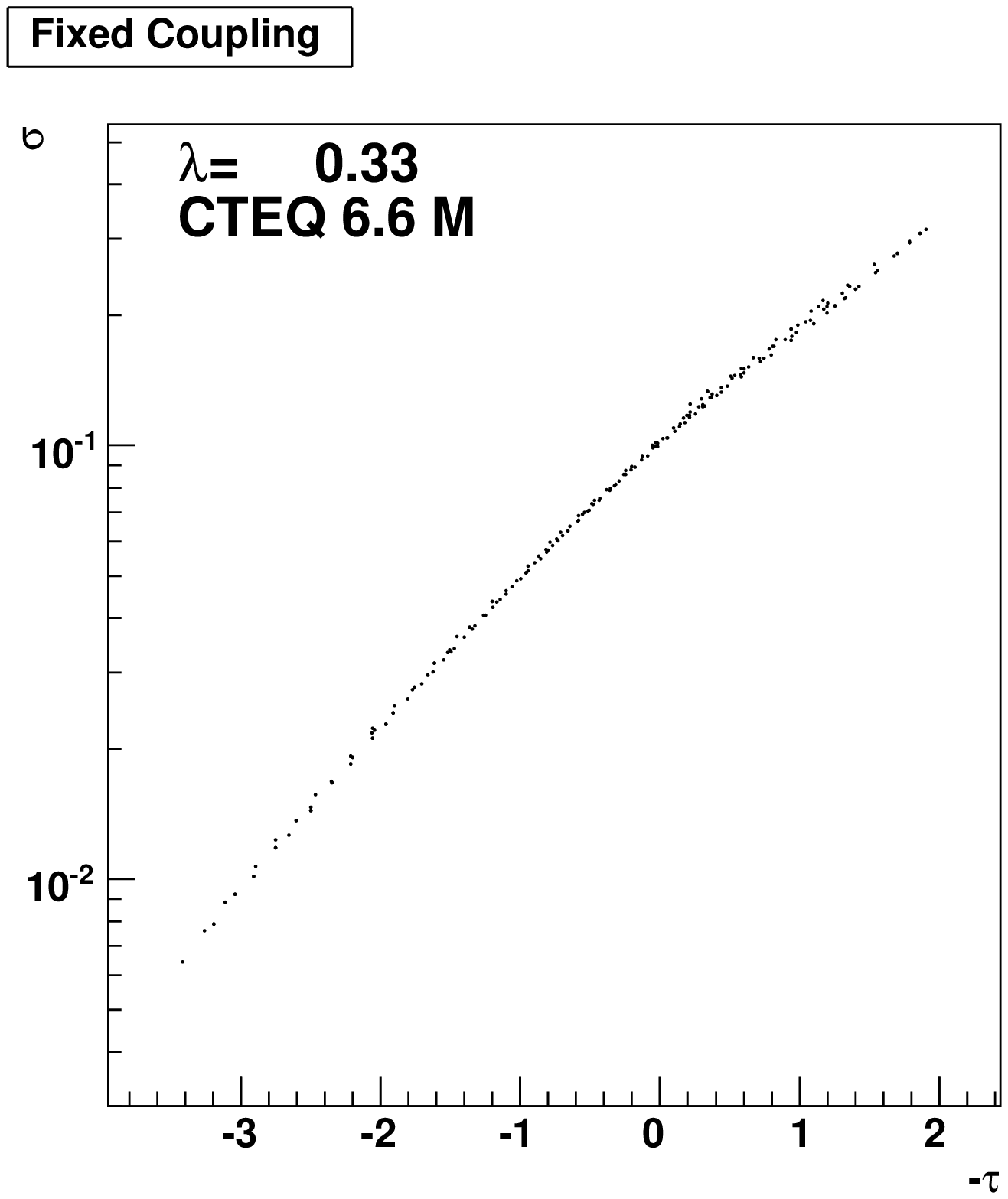,width=9.cm} &
\epsfig{file=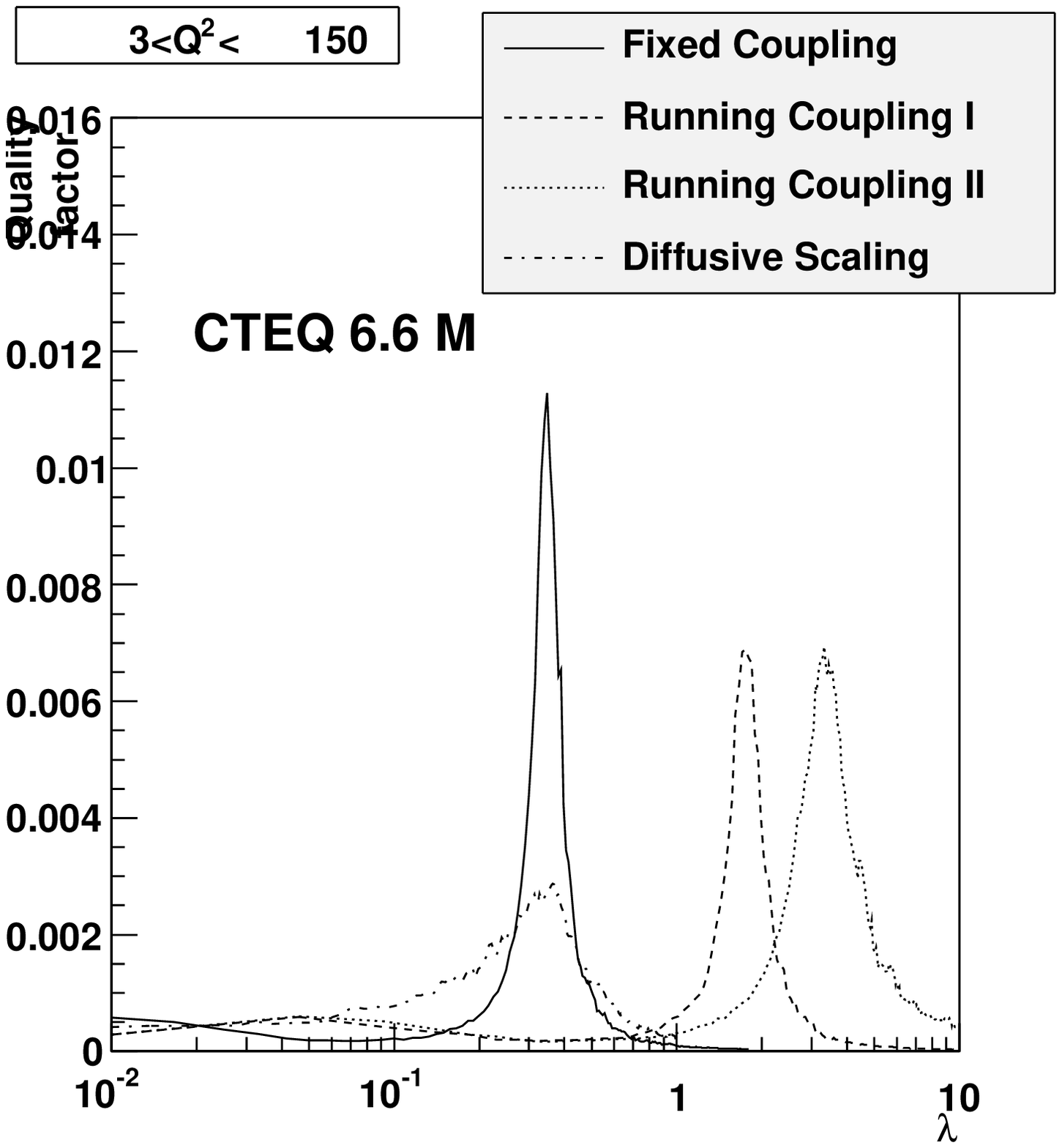,width=9.cm}
\end{tabular}
\caption{{\bf Different scalings for $F_2/Q^2$ measured data and the
CTEQ6.6M parametrisation for $Q^2>3$ GeV$^2$:} Upper curves:
Scaling curve for ``Fixed Coupling" (left) and $QF$ against $\lambda$ for
different scalings laws in data (right). Lower curves: same studies for
the CTEQ6.6M parametrisation.
Results are found similar for the different
CTEQ, GRV and MRST parametrisations and are similar to data.
}
\label{F2_dataparam}
\end{center}
\end{figure}

Table~\ref{f2_table} shows the fit results for all data and the parametrisations.
Similarly as in data, the MRST and GRV parametrisations give similar $QF$ for
fixed coupling, running coupling I and running couping II. On the contrary,
CTEQ favours fixed coupling.
All parametrisations disfavour diffusive scaling,
and all $\lambda$ values are consistent among different data sets. We can also see that
GRV leads to smaller values of $QF$ than MRST and CTEQ, but all parametrisations
lead to a good $QF$.
Scaling curves and fit results for experimental data and the CTEQ6.6M 
parametrisation, as an example,
are shown in figure~\ref{F2_dataparam}.

We note that the CTEQ, MRST and GRV parametrisations of the data
lead to good scalings, whereas they do not contain any saturation hypothesis.
One may wonder if such result comes from the chosen form of the parametrisation
at the initial scale $Q_0$, or if DGLAP evolution itself leads to a scaling in the HERA kinematical range.

\section{Fits to the charm component of DIS}

In a second part of the paper, we want to study the scaling properties
of the charm component of the structure function $F_2^c/Q^2$, both in data and 
in parametrisations.

\subsection{Fits to $F_2^c/Q^2$ in data and parametrisations}

First we test the scaling properties using experimental data.
The requirements on the kinematical domain remain the same as in the previous section.
Now the lower $Q^2>3$~GeV$^2$ cut also allows to remove a part of the charm mass effects.
We use the charm $F_2^c$ measurements from the H1, ZEUS and EMC experiments~\cite{charm}.
Only 25 data points lie in the kinematical region we want to analyse.

Since the statistics in the data is low, the fit results are not precise.
Nevertheless, they still lead to clear results that are comparable to $F_2$ fits.
Table III shows the results of the fits to the $F_2^c$ data
for different scaling laws.
The results are found similar between $F_2$ and $F_2^c$ (see
Fig. 2, top left plot).
All $\lambda$ parameters are similar except for Diffusive Scaling.
As in the case of the $F_2$ scaling analysis~\cite{usscaling}, Fixed Coupling, Running Coupling I
and Running Coupling II give similar values of $QF$, and
Diffusive Scaling is disfavoured.

It is now interesting to check whether the scaling properties are also
observed using the $F_2^c$ parametrisation of the data, as seen
in the previous section for the proton structure functions $F_2$. The results
are given in Table III and Fig. 2. Fig. 2 displays the values of the
$QF$ for Fixed Coupling and different parametrisations. The value of $\lambda$
favoured by the GRV and MRST parametrisations has a tendency to be larger
(about 1) than the value favoured in data (about 0.3). It is worth noticing
however that the data show a smaller peak in the $QF$ distribution around
1 as well which is disfavoured, and the MRST parametrisation another peak towards
0.3. The CTEQ parametrisation is definitely closer to the results in data
leading to a value of $\lambda$ close to 0.5 for the CTEQ6.6M NLO 
parametrisation, and to 0.4 when one introduces an additional sea-like intrinsic charm
contribution\footnote{CTEQ6.6C1 and CTEQ6.6C2 lead to the same
results as for CTEQ6.6M since the intrinsic charm contribution only appears
at high $x$ and does not affect our study performed for $x<10^{-2}$.}  
(CTEQ6.6C3 and CTEQ6.6C4). The results for the different scalings are
given in Table
III, and the conclusion remains unchanged. 

Let us now summarize what we obtained comparing data and simulation of $F_2^c$.
First of all, the value of $\lambda$ favoured in data is compatible with the one
obtained for $F_2$, DVCS... Of course the value of $\lambda$ shows a large error
bar since the number of data points is only 25, but it is quite striking that
the same value as for $F_2$ is found. Everything looks as if the charm quark was
behaving like any other light quark from the point of view of scaling. New data
on $F_2^c$ which will be published soon by the H1 and ZEUS collaborations are of
great interest. The MRST and GRV parametrisations favour different values of
$\lambda$ while the CTEQ ones are compatible with data, especially when an
instrinsic charm component is added. 

The next step of our study relies more in understanding in more detail the
differences between the parametrisations, since we recall that they lead to a good
description of $F_2^c$ itself.

\subsection{Comparison between different $F_2^c$ parametrisations}

Taking the same 25 points as in data is not sufficient
to investigate differences among the parametrisations.
In order to analyse the parametrisations in detail,
we choose to use the same $x$ and $Q^2$ as for $F_2$
which leads to 217 points.

The results for the 217 points are given in Table~\ref{cteq_table} for the CTEQ
parametrisation. For the sake of clarity,
we choose to perform the study for Fixed Coupling geometric scaling only ---
the differences among different scaling laws are not large and do not modify
the conclusion of the study.
We note that scaling is indeed obtained in the different CTEQ parametrisations,
and the value of
$\lambda$ which maximises $QF$ is compatible with the value measured in data.
The CTEQ6.6C4 parametrisation, which has a strong sea-like intrinsic
charm contribution, gives the $\lambda$ value that is closest to $\lambda=0.3$
found in data and 
leads to a much better $QF$ than the other parametrisations. It is also worth
noticing that the values of $\lambda$ are quite stable for different $Q^2$ cuts.

A new striking result appear with the MRST and GRV parametrisations when one
uses 217 points.
The MRST 2006 NNLO and the MRST 2004 NLO and NNLO parametrisations, as well as 
the GRV98 one,
do not show any scaling at all. 
No value of $\lambda$ that maximises $QF$ is found within an acceptable range.
We now study the behaviour of the MRST and GRV parametrisations after different
cuts on $Q^2$.
Increasing the cut to 5 GeV$^2$ does not help,
therefore we choose to study the effects of higher $Q^2$ cuts.
Table~\ref{mrst_table} gives the fit results in three different $Q^2$ ranges,
namely $[10; 150]$, $[15; 150]$ and $[25; 150]$~GeV$^2$.
Scaling properties are not observed using the MRST parametrisations
even in the
$Q^2$ range $[10; 150]$~GeV$^2$. The values of $QF$ are much smaller than the
ones obtained with CTEQ (see Table~\ref{mrst_table}). However,
geometric scaling appears at higher $Q^2$ 
and the value of $\lambda \sim 0.65$.
The situation for the GRV98 parametrisation is
found to be similar as for MRST, but leading to a higher value of $\lambda$.

As a conclusion to this second study, it looks like the MRST, GRV and CTEQ
parametrisations of $F_2^c$ show a different behaviour from the scaling point of view.
While the CTEQ parametrisation shows a scaling even at lower $Q^2$ (which seems
observed in data), the MRST parametrisation only shows scaling at higher $Q^2$.
Further more, the GRV parametrisation leads to a much higher value of $\lambda$
than the one favoured in data. This difference might be due to the different
forms of the gluon distribution at the initial scale $Q_0^2$ chosen by the CTEQ
and MRST groups.
It will be worth to test this further using new measurements of $F_2^c$ which
should be available soon.

Finally, let us mention that we have checked that substitution in the scaling 
formulae of $x$ by the rescaled variable $\chi$ (see Formula \eqref{ACOTchi}) used in the 
GM-VFNS has a negligible effect in the considered kinematical ranges and does 
not change our results and conclusions.

\begin{table}
\begin{center}
\begin{tabular}{|c||c|c|c|c|} \hline
data set & FC & RC I & RC II & DS \\
\hline\hline
data
   & $\lambda$=0.33 & $\lambda$=1.84 & $\lambda$=3.44 & $\lambda$=0.36 \\
   & $QF$=1.63 & $QF$=1.62 & $QF$=1.69 & $QF$=1.44 \\
\hline\hline
GRV98
   & $\lambda$=0.33 & $\lambda$=1.87 & $\lambda$=3.35 & $\lambda$=0.37 \\
   & $QF$=5.5 & $QF$=5.2 & $QF$=4.8 & $QF$=3.7 \\
\hline\hline
MRST 2004 NLO
   & $\lambda$=0.34 & $\lambda$=1.72 & $\lambda$=3.31 & $\lambda$=0.31 \\
   & $QF$=11.3 & $QF$=9.6 & $QF$=7.2 & $QF$=4.4 \\
\hline
MRST 2004 NNLO
   & $\lambda$=0.34 & $\lambda$=1.72 & $\lambda$=3.30 & $\lambda$=0.36 \\
   & $QF$=11.2 & $QF$=11.3 & $QF$=7.6 & $QF$=4.3 \\
\hline\hline
CTEQ 6.6 M
   & $\lambda$=0.35 & $\lambda$=1.84 & $\lambda$=3.40 & $\lambda$=0.38 \\
   & $QF$=12.4 & $QF$=6.6 & $QF$=6.4 & $QF$=2.7 \\
\hline
CTEQ 6.6 C1
   & $\lambda$=0.35 & $\lambda$=1.84 & $\lambda$=3.40 & $\lambda$=0.38 \\
   & $QF$=12.3 & $QF$=6.5 & $QF$=6.4 & $QF$=2.7 \\
\hline
CTEQ 6.6 C2
   & $\lambda$=0.35 & $\lambda$=1.84 & $\lambda$=3.40 & $\lambda$=0.37 \\
   & $QF$=12.1 & $QF$=6.4 & $QF$=6.3 & $QF$=2.7 \\
\hline
CTEQ 6.6 C3
   & $\lambda$=0.33 & $\lambda$=1.84 & $\lambda$=3.33 & $\lambda$=0.31 \\
   & $QF$=11.7 & $QF$=6.2 & $QF$=7.2 & $QF$=2.9 \\
\hline
CTEQ 6.6 C4
   & $\lambda$=0.33 & $\lambda$=1.66 & $\lambda$=3.11 & $\lambda$=0.38 \\
   & $QF$=13.1 & $QF$=9.0 & $QF$=6.9 & $QF$=2.4 \\
\hline\hline
\end{tabular}
\end{center}
\caption{\label{f2_table} Values of QF and $\lambda$ parameters for $F_2$ data and the
different $F_2$ parametrisations from CTEQ, GRV and MRST and for fixed coupling,
running coupling I, running coupling II and diffusive scalings. We note that
all scalings lead to a good QF except for diffusive scaling and that the values
of the $\lambda$ parameters are found similar in data and in the
parametrisations}
\end{table}

\begin{table}
\begin{center}
\begin{tabular}{|c|c|c|c|c|} \hline
data set & FC & RC I & RC II & DS \\
\hline\hline
data
   & $\lambda$=0.34 & $\lambda$=1.72 & $\lambda$=3.18 & $\lambda$=0.22 \\
   & $QF$=1.21 & $QF$=1.20 & $QF$=1.17 & $QF$=1.16 \\
\hline\hline
GRV98
   & $\lambda$=1.05 & $\lambda$=5.67 & $\lambda>$10 & $\lambda$=0.17 \\
   & $QF$=28.4 & $QF$=25.2 & & $QF$=9.9 \\
\hline\hline
MRST 2004 NLO
   & $\lambda$=0.96 & $\lambda$=5.24 & $\lambda>$10 & $\lambda$=0.16 \\
   & $QF$=10.0 & $QF$=7.8 & & $QF$=11.3 \\
\hline
MRST 2006 NNLO
   & $\lambda$=0.98 & $\lambda$=5.35 & $\lambda>$10 & $\lambda$=0.16 \\
   & $QF$=8.3 & $QF$=6.7 & & $QF$=9.9 \\
\hline\hline
CTEQ 6.6 M
   & $\lambda$=0.49 & $\lambda$=2.58 & $\lambda$=7.73 & $\lambda$=0.56 \\
   & $QF$=28.3 & $QF$=33.3 & $QF$=29.1 & $QF$=36.9 \\
\hline
CTEQ 6.6 C3
   & $\lambda$=0.48 & $\lambda$=2.42 & $\lambda$=6.03 & $\lambda$=0.51 \\
   & $QF$=46.1 & $QF$=48.7 & $QF$=35.2 & $QF$=46.5 \\
\hline
CTEQ 6.6 C4
   & $\lambda$=0.42 & $\lambda$=2.00 & $\lambda$=4.50 & $\lambda$=0.31 \\
   & $QF$=51.7 & $QF$=54.9 & $QF$=40.1 & $QF$=48.2 \\
\hline\hline
\end{tabular}
\end{center}
\caption{\label{charm_table} Values of QF and the $\lambda$ parameters using $F_2^c$ data measured
at HERA and the different $F_2^c$ parametrisations for fixed coupling,
running coupling I, running coupling II and diffusive scalings. We note that the
values of $\lambda$ are similar than the ones found with the proton structure
function $F_2$ and that diffusive scaling is disfavoured.}
\end{table}

\begin{table}
\begin{center}
\begin{tabular}{|c||c|c|c|c|c|} \hline
& CTEQ 6.6 M & CTEQ 6.6 C1 & CTEQ 6.6 C2 & CTEQ 6.6 C3 & CTEQ 6.6 C4 \\
\hline\hline
$Q^2>3$ &
    $\lambda$=0.70 & $\lambda$=0.70 & $\lambda$=0.68 & $\lambda$=0.56 & $\lambda$=0.44 \\
   & $QF$=0.8 & $QF$=0.8 & $QF$=0.8 & $QF$=3.5 & $QF$=14.6 \\
\hline
$Q^2>5$ &
    $\lambda$=0.61 & $\lambda$=0.61 & $\lambda$=0.61 & $\lambda$=0.52 & $\lambda$=0.42 \\
   &  $QF$=5.0 & $QF$=5.0 & $QF$=5.0 & $QF$=12.5 & $QF$=19.6 \\
\hline
$Q^2>10$ &
    $\lambda$=0.56 & $\lambda$=0.56 & $\lambda$=0.56 & $\lambda$=0.50 & $\lambda$=0.43 \\
   & $QF$=19.2 & $QF$=19.3 & $QF$=19.4 & $QF$=22.2 & $QF$=21.2 \\
\hline\hline
\end{tabular}
\end{center}
\caption{\label{cteq_table} Values of QF and of the $\lambda$ parameter for $F_2^c$ data and the different
CTEQ parametrisations for different $Q^2$ lower cuts for fixed coupling scaling.
We note that the scaling is observed in the CTEQ parametrisations and the value
of $\lambda$ is stable for different $Q^2$ cuts.}
\end{table}

\begin{table}
\begin{center}
\begin{tabular}{|c||c|c|c|c|} \hline
& MRST 2004 NLO & MRST 2004 NNLO & MRST 2006 NNLO& GRV98 \\
\hline\hline
$Q^2>10$ &
   $\lambda$=0.80 & $\lambda$=0.80 & $\lambda$=0.80 & $\lambda$=0.94 \\
   & $QF$=2.7 & $QF$=3.0 & $QF$=2.5 & $QF$=4.4 \\
\hline
$Q^2>15$ &
   $\lambda$=0.68 & $\lambda$=0.70 & $\lambda$=0.71 & $\lambda$=0.90 \\
   & $QF$=7.9 & $QF$=7.9 & $QF$=6.7 & $QF$=8.7 \\
\hline
$Q^2>25$ &
    $\lambda$=0.66 & $\lambda$=0.64 & $\lambda$=0.67 & $\lambda$=0.87 \\
   & $QF$=18.5 & $QF$=18.7 & $QF$=18.7 & $QF$=21.2 \\
\hline\hline
\end{tabular}
\end{center}
\caption{\label{mrst_table} Values of QF and of the $\lambda$ parameter for $F_2^c$ and the different
MRST and GRV parametrisations for different $Q^2$ lower cuts for fixed coupling scaling.
Scaling is not observed for a $Q^2$ cut lower than 10 GeV$^2$ in the
parametrisations.
We note that the quality of scaling at low $Q^2$ is
poor. For a higher $Q^2$ cut, scaling is observed but the value of the $\lambda$
parameter is different from the one in data for the GRV parametrisation.}
\end{table}

\begin{figure}[t]
\begin{center}
\begin{tabular}{cc}
\epsfig{file=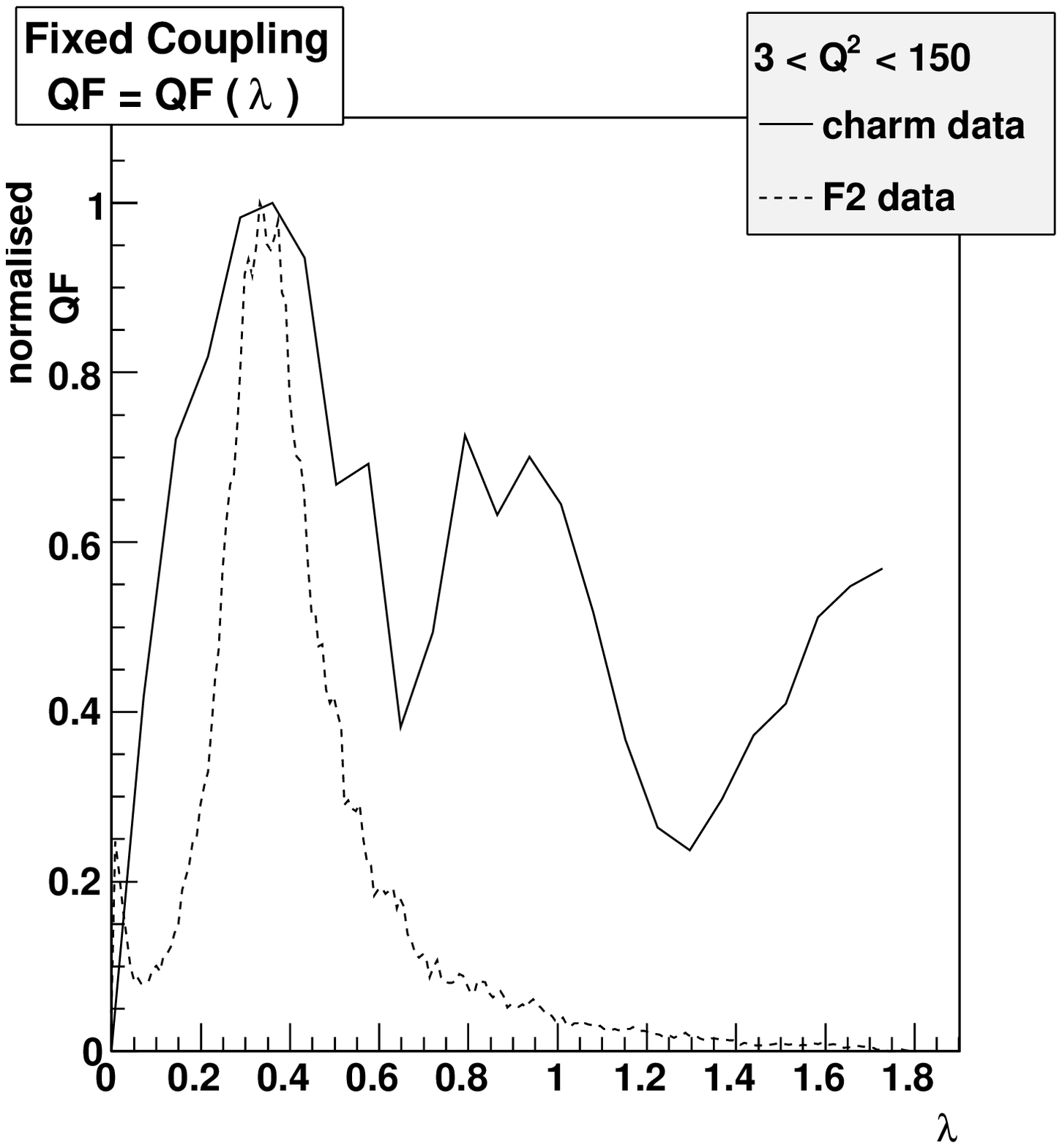,width=9.cm} &
\epsfig{file=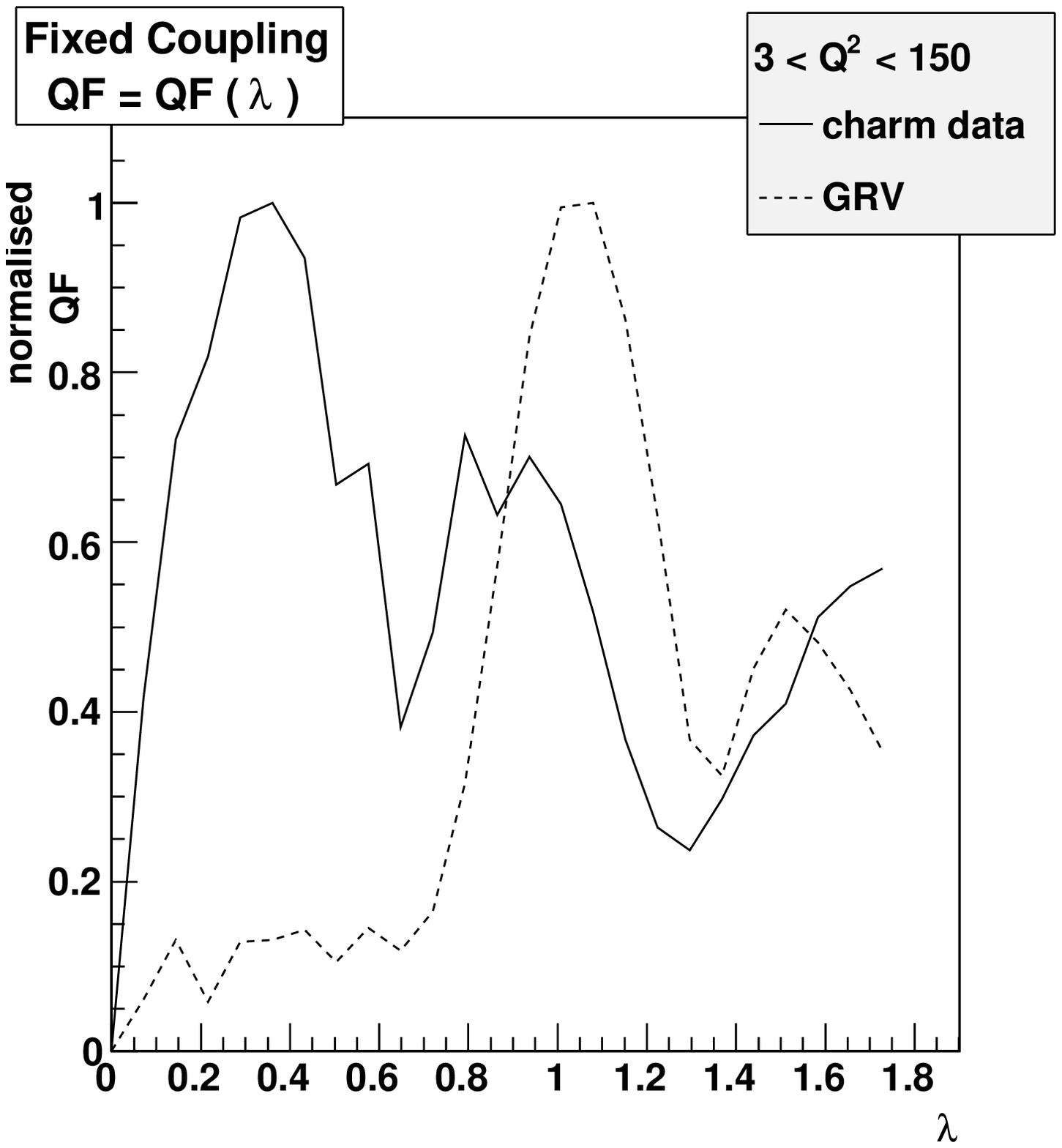,width=9.cm} \\
\epsfig{file=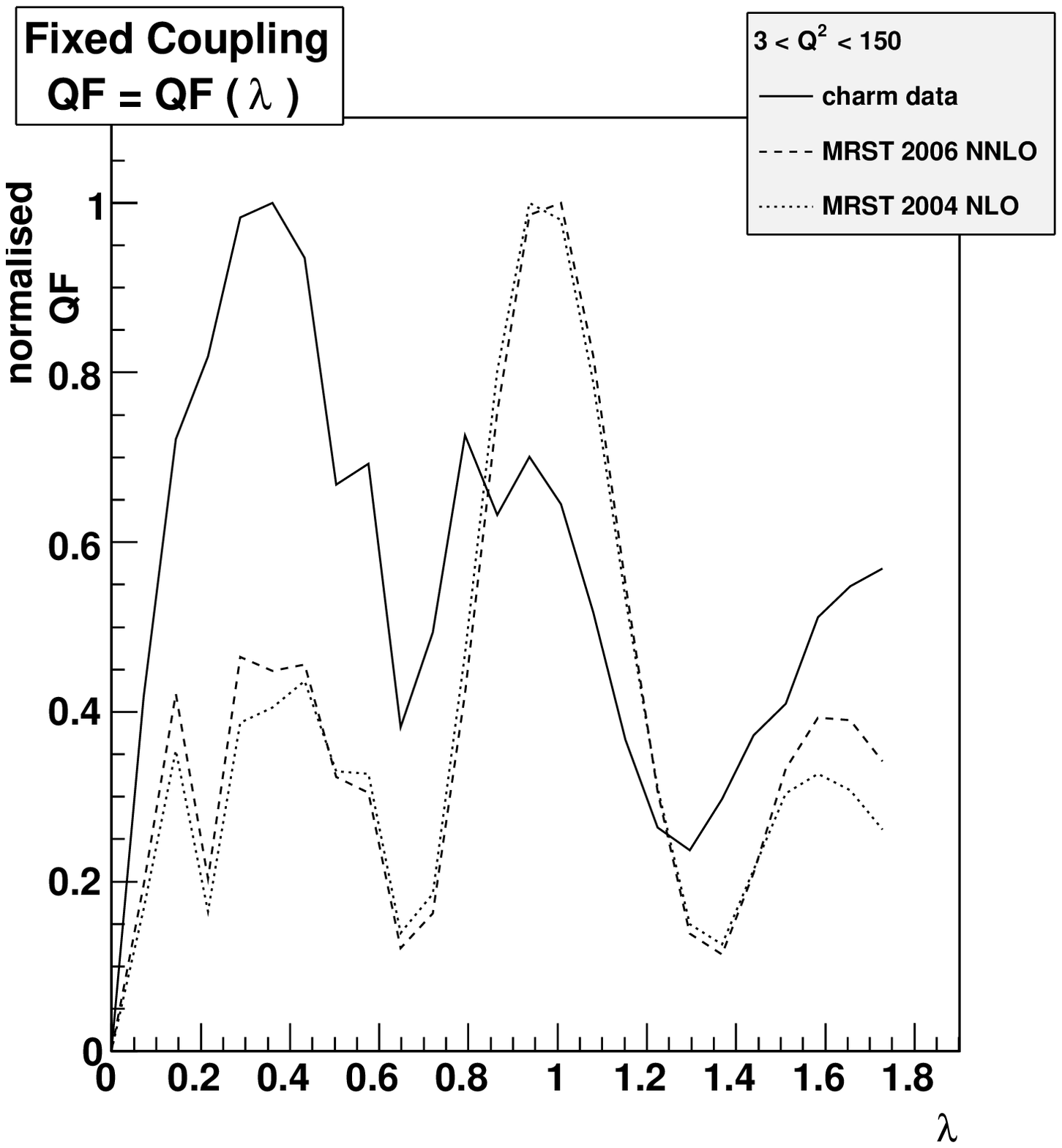,width=9.cm} &
\epsfig{file=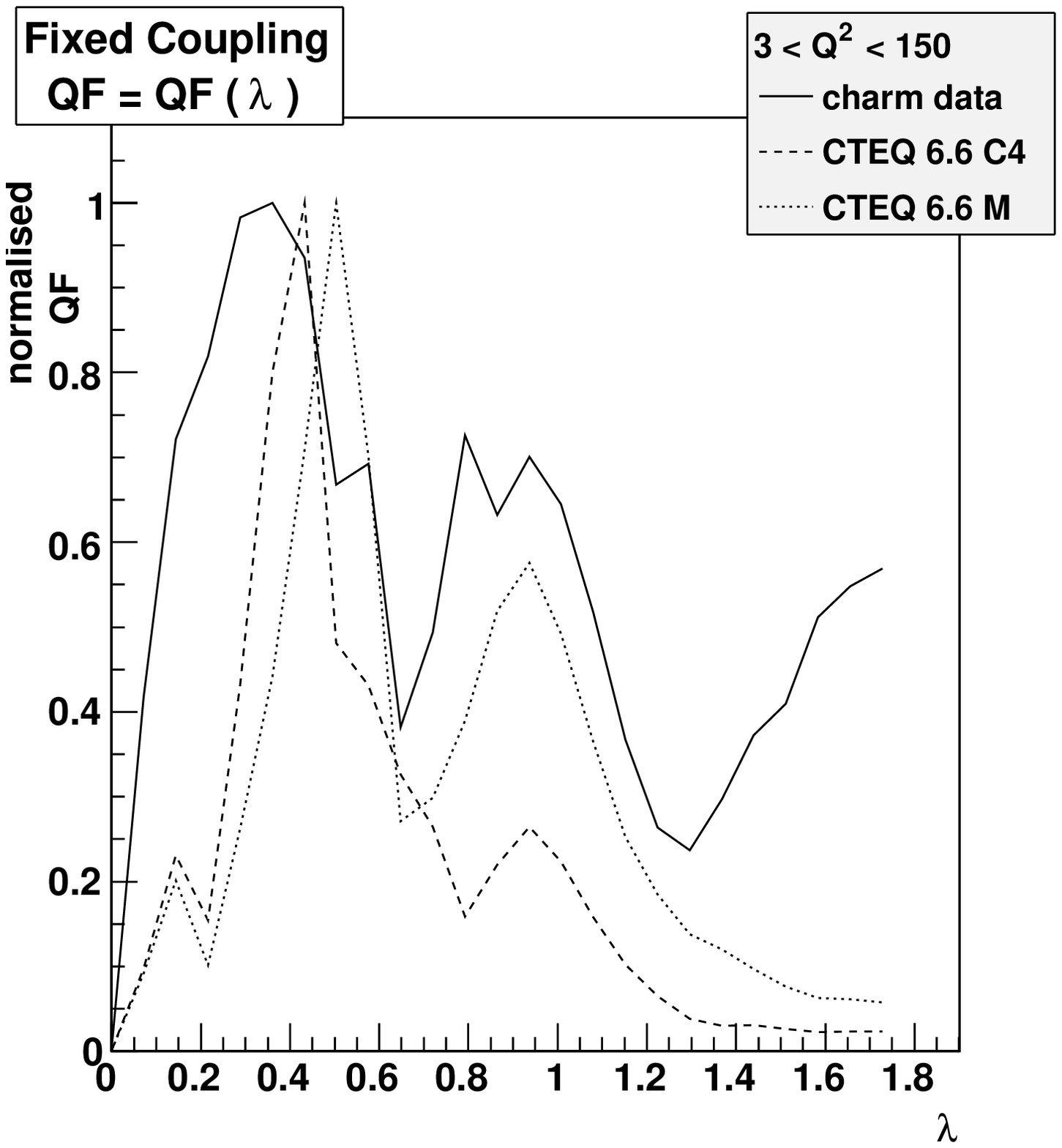,width=9.cm}
\end{tabular}
\caption{Quality factor as a function of $\lambda$ for Fixed Coupling using 
different data sets.
Quality factor peaks at the same value in $F_2$ and $F_2^c$ data.
Quality factor in the $F_2^c$ parametrisations peaks at higher values of 
$\lambda$ than in $F_2^c$ data.}
\label{QF25}
\end{center}
\end{figure}

%

\section{Conclusions}

We have investigated with the Quality Factor method the scaling properties of 
$F_2/Q^2$ and of its charm component $F_2^c/Q^2$, both in the data and in a 
selection of PDFs parametrisations. The parametrisations lead to the same 
scaling properties of $F_2/Q^2$ as in data for all types  of scaling 
variables suggested by gluon saturation theory. Thanks to the precision of the 
$F_2$ data, the global fits can indeed catch precisely the evolution of $F_2$. 
The $F_2^c$ HERA data contain few points and still show 
large error bars. The $QF$ method shows 
that the $F_2^c$ data favors precisely the same value of the parameter 
$\lambda$ for each type of scaling as the one found by the $QF $ fit on $F_2/Q^2$ 
and on DVCS data \cite{usscaling}.
The CTEQ parametrisation gives the closest scaling properties for $F_2^c/Q^2$
to the data, and the MRST and GRV parametrisations lead to favoured values of 
$\lambda$ higher than in data. 

To study further the differences between the parametrisations, we used 217
points, at the same $x$ and $Q^2$ values as for $F_2$. 
The parametrisations of the MRST and GRV groups have no scaling behavior 
at all if one includes points with $Q^2$ in the range $[3;10]$ GeV$^2$. 
Only the CTEQ (and especially the CTEQ6.6C4 parametrisation, which includes a 
strong sea-like intrinsic charm component) has a scaling behavior with a value 
of $\lambda$ close to data. Moreover, it is the only parametrisation which 
leads to a very good scaling of $F_2^c/Q^2$ for $Q^2$ down to $3$ GeV$^2$. 
The fact that the
$F_2/Q^2$ and $F_2^c/Q^2$ data shows the same scaling 
properties seems to indicate that the behavior of the charm quark distribution 
in the proton is closer to light sea quarks in data than in most 
parametrisations.

Of course, our study rely to a large extent on the $QF$ analysis of the $F_2^c$ 
data which are not very precise. Hence, new $F_2^c$ data with improved 
statistics would be welcome, in order to check accurately if the scaling 
properties of $F_2/Q^2$ and $F_2^c/Q^2$ are indeed the same.
However, as the coincidence for the optimal value of $\lambda$ for 
$F_2^c/Q^2$, $F_2/Q^2$ and DVCS is very striking, let us consider that 
there could be a deep reason for that, and discuss possible explanations.

Since the behavior of the GRV98, which is based on a FFNS, and of recent MRST 
parametrisations based on GM-VFNS is similar, we conclude that the type of 
scheme for heavy flavors does not change much our discussion. Similarly, 
the order (NLO or NNLO) of the global fit seems not to be essential for the 
scaling properties. Thus, the similarity of the scaling properties of 
$F_2/Q^2$ and $F_2^c/Q^2$ in the data seems to result from a theoretical 
ingredient at low $Q^2$ or low $x$, which is missing in all of the usual 
global fits.

The parametrisation which gives the best scaling properties and is the 
closest to data is CTEQ6.6C4. It could be tempting to say that our 
study gives hints for a strong sea-like intrinsic charm component in the 
proton. However, this is not the only possible explanation. For example, 
low $x$ resummations and/or saturation effects would lead to a larger and more 
stable gluon distribution at low $x$ and low $Q^2$ in a global fit. 
This would enhance the charm contribution produced by boson gluon 
fusion in that kinematical range without the need of intrinsic charm. 
Moreover, in that scenario, the light and heavy sea quark distributions 
would be both driven by the gluon distribution rather than by their initial 
conditions, and thus would naturally behave in a similar way.

\begin{acknowledgments}
The authors thank Robi Peschanski for useful discussions, and Jon Pumplin for suggesting the study of the $F_2^c$ scaling properties.
\end{acknowledgments}

\end{document}